\documentstyle[preprint,aps]{revtex}

\begin{document}
\draft
\title{
Nonlinear Magnetization in Superconductors with s + d 
Ordering }
\author{J.J. Betouras and Robert Joynt}
\address{
Department of Physics and Applied Superconductivity Center\\
University of Wisconsin-Madison\\
Madison, Wisconsin 53706\\}
\date{\today}
\maketitle

\begin{abstract}

The nonlinear magnetization is considered within the Ginzburg-Landau 
theoretical framework, in the Meissner regime. A calculational method in the 
case of conventional superconductors (one order parameter) is developed and 
this method is extended for the case of two order parameters (s+d mixing). It 
is confirmed that corrections to the penetration depth, in the mean field 
analysis, are of the order of ${H_0}^2$ where $H_0$ is the applied field. We 
analyze carefully the possible solutions which lead to different scenarios in 
the physics of the symmetry of the order parameter. The anisotropy in the 
penetration depth is calculated  and the temperature dependence of the 
magnetization is extracted. We discuss the relevant experimental results 
in the light of these calculations.

\end{abstract}
\narrowtext
\newpage

\section{INTRODUCTION}
\label{sec:level1}

The symmetry of the order parameter in high-T$_{c}$ superconductors
is an important and vital question. Knowledge of the symmetry of the 
superconducting gap function adds information regarding the microscopic 
mechanisms and reveals novel phenomena as well. An important contribution to 
this effort comes from one particular experimental technique, the measurement 
of the penetration depth or the magnetization when magnetic field is applied.
The dependence of the penetration depth on the applied field (for small 
values of the field) and on the temperature is able to provide some 
conclusions on the order parameter symmetry. 
The quantity which is most easily measured is the
deviation of the penetration depth from its zero-temperture,
zero-field value: $\Delta \lambda_{ab}(H,T) = 
\lambda_{ab}(H,T) - \lambda_{ab}(0,0)$.  The indices indicate an 
in-plane penetration depth. The measurements that have been performed on 
different materials do not give a completely clear picture.
Measurements on $YBa_2Cu_3O_y$ crystals by Sridhar $et$ $al.$ \cite{shridhar}
result in a quadratic field-dependent penetration depth below 
$H_{c1}$ and linear dependence above $H_{c1}$. On the other hand 
$Bi_2Sr_2CaCu_2O_y$ (BSCCO) shows a quadratic behavior in a region of values 
of magnetic field much smaller than $H_{c1}$ and linear dependence in the 
remaining region \cite{maedaprl}. The temperature dependence shows a
very interesting behavior as well. Hardy $et$ $al.$ observed 
linear  T-dependence at low temperatures in YBCO \cite{hardy} and 
Bonn $et$ $al.$ observed a crossover from linear-T to $T^2$ upon Zn doping 
\cite{bonn}. Also a $T^2$ behavior has been observed on BSCCO by Maeda 
$et$ $al.$ \cite{maedaprb}.
The linear field-dependence is predicted by the 
d-wave scenario, according to Xu $et$ $al.$ \cite{sauls}.

At this moment the question of the microscopic 
mechanism is still lacking a clear consensus,
though most work now concentrates on the magnetic mechanism (antiferromagnetic 
fluctuations in $CuO_2$ planes) as the first candidate for the d-wave 
component. 
Experimental results support an order parameter of
predominantly d-wave symmetry, as reviewed in Refs.\ \cite{vanharlingen}
and \cite{sigr}.  However, there are possible indications of an admixture
of s-wave in overdoped and electron-doped materials, as 
reviewed in Ref.\ \cite{kj}.  This latter possibility is consistent with
the magnetic hypothesis. 

A signature of s-d mixing is the spontaneous appearance 
of orthorhombic-like anisotropies in the superconducting state.  In the
penetration depth this would mean $\lambda_x \neq \lambda_y$.
In a system such as YBCO with an orthorhombic crystal structure, it is 
necessary to somehow disentangle this spontaneous anisotropy from the 
crystalline anisotropy.  This paper attempts to do this for one particular 
measurable quantity, the temperature- and field-dependent penetration depth.

Experiments in YBCO have indeed shown a large difference in the penetration 
depth along the a axis ($\lambda_a$) and b axis ($\lambda_b$), which has been 
attributed to the orthorhombic distortion \cite{dynes,basov}. This
distortion can be taken into account in the construction of the free energy.
In Ref. \cite{carbotte1} the anisotropy at zero temperature has been calculated
in a microscopic model where the pairing is of d-wave type and using a single
tight-binding band with different hopping parameters in the a and b
directions. The point of view is that the chains accomodate a significant part
of the condensate and this has to be taken into account. 
Also in Ref. \cite{carbotte2} it was found that the chains cannot become 
superconducting by proximity effect. The microscopic theory of the non-linear
Meissner effect within a purely one-component d-wave scenario has been 
formulated by Xu $et$ $al.$ \cite{sauls}.  Their emphasis was on low 
temperature behavior, while our Ginzburg-Landau approach is more suited to 
temperatures neat T$_c$.

In this paper, the  framework of Ginzburg-Landau (G.L.) theory is used and 
this mean field theory allows us to do calculations taking into consideration 
the possible symmetry of the order parameter and constructing the appropriate 
free energy functional. In this manner the specific microscopic mechanism is 
avoided and the thermodynamic properties can be extracted. The attention is 
concentrated in the Meissner phase, where the presence of vortices can be 
neglected and the field can be considered weak. The solution of the 
G.L. equations beyond the London approximation offers the 
dependence of the penetration depth and the magnetization on low applied 
fields and also the temperature. In mean field theory then the dependence on 
the field is quadratic. The corrections of the order parameter can be 
calculated analytically as well.
 
The plan of the paper is as follows. In Sec.\ II, the G.L. equations 
are solved in the case of conventional superconductors in the Meissner phase, 
so the calculational method is illustrated.
We believe it is really worth presenting these calculations, since we
were unable to find any detailed calculation in the literature. In Sec.\ III, 
the method is extended to the case of the mixed symmetry in the order 
parameter ( s + d). The possible solutions are considered and we discuss 
the implication of each one. In general the situation where the d-wave 
is dominant is adopted and the role of the s-wave component is illustrated.
This s-symmetry part appears partly because of the perturbation
due to the field but a nonzero value even in the absence of external field
can be present. The anisotropy in the penetration length along the a and 
b axes can be calculated easily since the orthorhombic 
distortion is considered in the development of the free energy functional.
In Sec.\ IV  we conclude and discuss the results in connection to relevant 
experimental data. Also in the Appendix the full set of equations 
are given and the implication of the different field directions are considered.

\section{ONE ORDER PARAMETER - CONVENTIONAL SUPERCONDUCTORS}
\label{sec:level2}

For the case of the conventional (s-wave) superconductors, the
G.L. free energy is written:
\begin{eqnarray}
F = \int \left[\alpha|\Psi(x)|^2+\beta|\Psi(x)|^4+\gamma|\vec{\Pi}\Psi|^2 +
\frac{h^2}{8 \pi} \right] d^3 x
\label{glfunc1}
\end{eqnarray}

where  $\vec{\Pi} = -i \hbar \vec{\nabla} - \frac{2e}{c}
\vec{A}$  and $ \vec{B} = \vec{\nabla} \times \vec{A} $.
The above free-energy is minimized  with respect to $\Psi^{*}$ and $\vec{A}$.
$\Psi$ is written as $\Psi = |\Psi|\;e^{i\phi} $, where 
$\phi$ is the phase of the order parameter.
Then the equations to be solved are :
   
\begin{eqnarray}
\gamma \left[-\hbar^2\vec{\Pi}^2|\Psi|+  (\hbar\vec{\nabla}\phi-\frac{2e}{c}
\vec{A})^2 |\Psi|\right] + \alpha\Psi + 2\beta|\Psi|^2\Psi = 0
\label{one}\\
\frac{c}{4\pi} \vec{\nabla} \times \vec {\nabla}\times\vec{A}=
4\gamma e |\Psi|^2(\hbar\vec{\nabla}\phi-\frac{2e}{c}\vec{A})
\label{two}
\end{eqnarray}

In the London approximation $|\Psi|$ is considered spatially constant, 
unaltered by the applied field. From Eq. (\ref{two}) above the penetration 
depth is found to be $\lambda = \sqrt{ (-c^2 \beta/16 \pi \gamma \alpha  
e)}$. In the calculations, the applied magnetic field $\vec{H_0}$ is taken
parallel to the z-axis, the superconductor occupies the half space $x>0$
and the most convenient gauge in which to work is the one with $A_y=A_z=0$. 
Then the phase $\phi$ and the vector potential $A_x$ have x,y- dependence only.
Since the field and the supercurrent are expected to have only x-dependence 
(translational invariance in the $\hat{y}$ direction) $\phi$ and $A_x$ can be 
written  as : $\phi= y \;g(x)$ and $A_x = y\;a(x)$. $|\Psi|$ is expected to have
x-dependence as well.  After these substitutions (also
for convenience  $|\Psi| \equiv f$) the above equations become :

\begin{eqnarray}
\gamma(-\hbar^2 \frac{d^2f}{dx^2} + \hbar^2 g^2 f) + 
2 \beta f^3 + \alpha f =0 
\label{three}\\
\frac{dg}{dx} = \frac{2e}{\hbar c}\; a(x) 
\label{four}\\
\frac{da}{dx} = \frac{8\pi e}{c} f^2 \hbar g 
\label{five}
\end{eqnarray}

Eq. (\ref{three})/(\ref{four}) above is the real/imaginary part of 
Eq. (\ref{one})
respectively. The boundary conditions are: $a(x=0) = H_0$, $a(x=\infty)=0$,
$g'(x=0)= (\hbar c/2e) H_0$, $g'(x=\infty)=0$, $f(x=\infty) 
=f_0=\sqrt{-\alpha/2\beta}$ and  $f'(x=0) = 0$  \cite {degennes}. 
It is obvious then that $g \propto H_0$ and the second 
term in the l.h.s. 
of Eq. (\ref{three}) above serves as a term containing 
the small parameter of the problem. 
Then perturbation theory can be used to obtain the solution in small fields 
(small compared to the thermodynamic field $H_c$). We consider solutions of 
the form:

\begin{eqnarray}
f = f_0 + H_0^2 \;f_1 + H_0^4 \;f_2 + ...
\end{eqnarray}

In this case, if the above ansatz substitutes f in Eq. (\ref{three}) and
terms with same power in $H_0$ are gathered ($\tilde{g} \equiv g/H_0$)
we get:
\begin{eqnarray}
-\gamma\hbar^2 \frac{d^2f_0}{dx^2}+ 2\beta f_0^3+\alpha f_0 &=& 0 \\
-\gamma\hbar^2 \frac{d^2f_1}{dx^2} + 6\beta f_0^2f_1+\alpha f _1 &=&
-\gamma\hbar^2g^2f_0 \\
\frac{\hbar c}{8\pi e} \frac{d^2\tilde{g}}{dx^2}&=&
\frac{2e\hbar}{c}(f_0^2+2H_0^2f_1f_2) \; \tilde{g} 
\end{eqnarray}          

The last equation comes from the combination of Eqs. (\ref{four}) and 
(\ref{five}).
The above expansion is correct, since the equations corresponding to greater 
powers of $H_0$ are such that convergence is guaranteed. Therefore the above 
method works and leads to correct results. Then the largest correction due to 
the field will have an $H_0^2$ dependence. The solution for $f_0$ is the 
unperturbed part $\sqrt{-\alpha/2\beta}$. The solution for $f_1$ then can be 
found easily and the form of the order parameter becomes ($H_c \equiv 
\frac{\hbar c} {\sqrt{2}\xi \lambda e }$)
\begin{eqnarray}
|\Psi| = \sqrt{-\alpha /2\beta}\left\{ 1 + (\frac{H_0}{H_c})^2 
\frac{2}{(\frac{4 \xi^2}{\lambda^2} - \frac{1}{2}) }
\left[-\frac{2 \sqrt{2} \xi} {\lambda}  \exp(-x/\sqrt{2} \xi) 
+ \exp(-2x/\lambda) \right] \right\}.
\end{eqnarray}

An interesting immediate observation  is that besides the term with the 
exponential dependence on the coherence length there is another one which 
decays with the characteristic length of  $\lambda/2$. The above values of  
$\xi$ and $\lambda$ are the zero-field ones. In the case  of a superconductor 
with $\xi=\lambda/2 \sqrt{2}$ the divergent part in the denominator is 
cancelled by the numerator. Also, the nonlinear correction $f_1$ is negative, 
independent of the type of the superconductor, something which is expected 
since the magnetic field acts as a pair-breaking mechanism.

The next step is to substitute the above solution in the equation
of $\tilde{g}$ and solve it by the same method.
We define the effective penetration depth as follows:
\begin{equation}
\lambda_{eff} =  \frac{1}{H_0} \int_{0}^{\infty} a(x) dx
\end{equation}
By performing the integration the effective penetration depth
becomes :
\begin {equation}
\lambda_{eff} = \lambda(H=0) [1 + (\frac{H_0}{H_c})^2 \
\frac{\kappa (\kappa + 4 \sqrt{2})}{(\kappa + 2 \sqrt{2})^2} ]
\end{equation}

The penetration field is an increasing function of the field and has a 
quadratic dependence on it.  It is somewhat remarkable that this
expression does not appear anywhere in the published literature, as far as we 
are aware.  It holds for every value of 
$\kappa=\lambda(H=0)/\xi(H=0)$, and is therefore a 
generalization of a formula
given in Ref.\ \cite{shridhar}:
\begin{equation}
\delta \lambda/ \lambda = 3/4 \times (H_0/H_c)^2, 
\end{equation}
which holds in the limit $\kappa \rightarrow \infty $.
The finite $\kappa$ correction is due to the 
part of the order parameter that decays with 
the characteristic length of $\lambda/2$. 

\section{ s+d ORDER PARAMETER SYMMETRY}
\label{sec:level3}

Recent experimental evidence and also theoretical calculations \cite{ben}
suggest that a mixed-symmetry order parameter is a possible candidate to 
explain several features observed in experiments on high-$T_c$ materials in
the absence of a field. However, even if there is no s-wave component in
zero field, an s-wave component is always formed in the vicinity of 
vortex cores or due to induced currents or due to surface or impurity effects
or, finally, as a result of the orthorhombic distortion in these materials.
Within the Ginzburg-Landau theoretical framework, the free-energy 
functional that takes into account both the symmetry of the material and the 
order parameter (but with fourth-order derivatives neglected) 
takes the form \cite{bob} :

\begin{eqnarray}
F & = & \int d^3 x \left\{ \alpha_d |\Psi_d|^2+\beta_2 |\Psi_d|^4+
\gamma_d |\vec{\Pi}\Psi_d|^2+\alpha_s |\Psi_s|^2+ \beta_1
|\Psi_s|^4 + \gamma_s |\vec{\Pi}\Psi_s|^2 +  
\beta_3 |\Psi_s|^2 |\Psi_d|^2 \right.   \\
& & + \left. \beta_4 ( {\Psi_s}^{*2} {\Psi_d}^2 + {\Psi_d}^{*2} {\Psi_s}^2)
+ \gamma_v [ (\Pi_y \Psi_s)^* (\Pi_y \Psi_d) - 
(\Pi_x \Psi_s)^* (\Pi_x \Psi_d) + c.c. ] + \frac{h^2}{8\pi} \right\}.
\label{glfunc2} 
\end{eqnarray} 

The orthorhombic distortion has been taken into account, through the
mixed-gradient term (and a perturbative bilinear on the two components term).
These terms may serve as a ``source'' for the s-component 
(it's one of the two ``scenaria'' that come from the possible solutions).
The mixed-gradient term distinguishes a-axis from b-axis as well
(c-axis is along the z-direction).
The above functional can be used to derive the anisotropy in the penetration 
depth that is observed. The same G.L. functional has been extensively studied 
recently for the case of vortices  and actually has been derived in the 
weak-coupling limit for both continuous and lattice
Hamiltonians \cite{ting,kallinprl,kallinprb,ichiokaprb}. 
The geometry will be the same (Meissner-geometry) and the difference is that 
now penetration along both x and y directions have to be considered separately.
Another difference comes from the two-dimensional nature of the free energy 
functional. The orientation of the applied field produces somewhat different 
results so one has to consider both the ``in-plane'' case and the case where 
the field is along the z-direction.

\subsection{London Approximation}

After minimization Eq. \ref{glfunc2} with respect to ${\Psi_s}^*$, 
${\Psi_d}^*$ and $\vec{A}$ the Euler-Lagrange equations to be solved are:

\begin{eqnarray}
(\gamma_d {\Pi}^2 + \alpha_d) \Psi_d &+& \gamma_v ({\Pi_y}^2 - {\Pi_x}^2)
\Psi_s + 2\beta_2 |\Psi_d|^2 \Psi_d + \beta_3 |\Psi_s|^2 \Psi_s +
2 \beta_4 {\Psi_s}^2 {\Psi_d}^* = 0 
\label{six}\\
(\gamma_s {\Pi}^2 + \alpha_s) \Psi_s &+& \gamma_v ({\Pi_y}^2 - {\Pi_x}^2)
\Psi_d + 2\beta_1 |\Psi_s|^2 \Psi_s + \beta_3 |\Psi_d|^2 \Psi_d +
2 \beta_4 {\Psi_d}^2 {\Psi_s}^* = 0 
\label{seven}\\
\nonumber
\frac{c}{4 \pi} \vec{\nabla} \times \vec{\nabla} \times \vec{A} &=&
\frac{e^*c} {\hbar} \left\{ \gamma_d {\Psi_d}^* \vec{\Pi} \Psi_d \right. +
\gamma_s {\Psi_s}^* \vec{\Pi} \Psi_d + \gamma_v ( {\Psi_s}^*
(\hat{y} \Pi_y - \hat{x} \Pi_x) \Psi_d + \nonumber \\
&& \left. {\Psi_d}^* (\hat{y} \Pi_y -
\hat{x} \Pi_x ) \Psi_s 
+ c.c. \right\} 
\label{eight}
\end{eqnarray}

From the form of the above two first equations it's easy to verify that
the only possibility for the {\it relative} phase $\phi$ of the two
components is 0 or $\pi$, therefore only d$\pm $s states are the starting point
of our analysis. Physically d+s and d-s are equivalent and the
system spontaneously chooses one of these states. The two axes $\hat{x}$
and $\hat{y}$ are identified with the crystallographic a and b in order to
make connections with the experiments. 
The applied field is along the $\hat{z}$-direction 
and its spatial variation is along the $\hat{x}$-direction. The boundary
conditions are:

\begin{eqnarray}
\nonumber
|\Psi_s(x=\infty)| \equiv f_{s0} = const. \qquad
|\Psi_d(x=\infty)| \equiv f_{d0} = const. \qquad
\vec{A} (x=\infty) = 0 \\
\nonumber
[\gamma_d \vec{\Pi} \Psi_d + \gamma_v (\hat{y} {\Pi}_y - \hat{x} {\Pi}_x)
\Psi_s ] \cdot \hat{x} = 0   \qquad
[\gamma_s \vec{\Pi} \Psi_s + \gamma_v (\hat{y} {\Pi}_y - \hat{x} {\Pi}_x)
\Psi_d ] \cdot \hat{x} = 0  
\end{eqnarray}

$|\Psi_{s0}| \equiv f_{s0}$ and $|\Psi_{d0}| \equiv f_{d0}$ are the bulk values
of the two order parameters, without the applied field.
If the London approximation is made ($|\Psi_d|,|\Psi_s|$ spatially constant) 
the penetration depth along the two axes can be found  to be:

\begin{eqnarray}
\lambda_x = [\frac{c^2}{4 \pi e}  \frac{1}{(\gamma_d |\Psi_d|^2+
 \gamma_s |\Psi_s|^2 + \gamma_v |\Psi_s| |\Psi_d)|} ]^{1/2} \\
\lambda_y = [\frac{c^2}{4 \pi e} \frac{1}{(\gamma_d |\Psi_d|^2 +
 \gamma_s |\Psi_s|^2 - \gamma_v |\Psi_s| |\Psi_d|)} ]^{1/2}        
\end{eqnarray}

The above formulas give a first estimate how it is possible
to obtain the difference in the penetration depth from
the G.L. theory consistent with the experimental measurements \cite{dynes}.
The ratio $\lambda_x/\lambda_y$ is $1 - \epsilon$ where $\epsilon$ is the
quantity $\frac{\gamma_v \; |\Psi_s|}{\gamma_d \; |\Psi_d|} + 
O( (\gamma_v \; \frac{|\Psi_s|}{|\Psi_d|} )^2 )$.
In Fig. \ref{fig1} this ratio which measures the anisotropy has been plotted 
as function of temperature and,
as we see close to  $T_s$ where the s-component grows as we move from the
higher temperatures towards the lower ones, the anisotropy becomes stronger.
Since the existence of the s-component enhances the anisotropy, close to T$_c$
the anisotropy vanishes. 

What this calculation demonstrates is the fact that the anisotropy
can shhow a strong signature of s-d mixing if the ratio of the
two components is temperature-dependent.  The photoemission 
measurements on BSCCO-2212 \cite{photo1} have been interpreted
as showing just such a dependence \cite{jbrj}.  
This interpretation can therefore be 
checked by magnetization measurements, if careful attention is paid to 
anisotropy.

\subsection{Beyond the London Approximation}

For the calculations beyond the London Approximation the method of
Sec. II can be followed closely. From the nature of the equations there are 
two points of view in obtaining  the solutions, with substantial difference
in the physics of the problem. The first point of view, or the first physical
``scenario'' is the one starting from the assumption that $\alpha_s \geq 0$,
so that the zeroth order 
(in the magnetic field) value of the magnitude of the  $\Psi_s \equiv f_{s0}$ 
is 0. In other words, the onset of the s-component of the order parameter is 
caused by the application of the magnetic field  and the induced currents and
can be viewed as a perturbation to a robust d-wave superconductor,
or a small ``transformation'' from d to d+s, close to the boundary.
There are three characteristic length scales in the problem, 
the two coherence lengths $\xi_d$ and $\xi_s$ that characterize the spatial
change of the two distinct order parameters and the penetration depth
$\lambda$ which is the characteristic length of the electromagnetic changes. 

The second physical picture is when $\alpha_s < 0$ and a nonzero value of 
$f_{s0}$ at zero current is considered.
Then the superconductivity is truly two-channel and the existence of the 
s-wave part has to do either with the pairing mechanism or with the departure
from the tetragonal symmetry of the lattice and not with the applied magnetic 
field which at most modifies the form of the order parameter close to the 
boundary. In this picture there are four different characteristic length 
scales, the two coherence lengths $\xi_d$, $\xi_s$ and the two penetration 
depths  $\lambda_d$ and $\lambda_s$ (combined to one $\lambda$) :

\begin{equation}
\nonumber
\frac{1}{ {\lambda_{x,y} }^2 } = \frac{1}{ {\lambda_d}^2 } + 
\frac{1}{ {\lambda_s}^2 } \pm  \frac{\gamma_v}{ \sqrt{\gamma_d \; \gamma_s} } 
\frac{1}{\lambda_d \; \lambda_s}
\end{equation}

\subsubsection{Zero $|\Psi_s|$ at zero applied field}

In this case there are the following consequences :
(i) The onset of s-wave is due to the mixed gradient term and the value of
the order parameter  $|\Psi_s| \propto  \gamma_v \times {H_0}^2$.
(ii) Since this mixed gradient term is responsible for the s-wave it is
clear that when the y-dependence is examined alone one gets $|d+s|$ state,
on the other hand when the x-dependence is examined one gets $|d-s|$.
(iii) According to (i) and (ii) anisotropy $cannot$ be observed in
penetration depth due to term proportional to $\gamma_v$ in Eq.[15] and [16].
This term finally gets proportional to ${\gamma_v}^2$ and the sign difference
cancels out.
(iv) The only way to get the anisotropy in penetration depth is to
consider different values for the ``masses'' in different directions. Namely
$\gamma_{dx}$ and $\gamma_{dy}$ and the same for $\gamma_s$ possibly.
(v) Due to the above observations  the quantity $\lambda_x/\lambda_y$ is 
temperature- $independent$.

The two coherence lengths are given by the equations: 

\begin{eqnarray}
\xi_s = \sqrt{ \frac{{\hbar}^2 \gamma_s}
{-2 \alpha_s +2 \beta_3 ' {f_{d0}}^2} }  \qquad  \qquad
\xi_d = \sqrt{  -\frac{{\hbar}^2 \gamma_d}  {4 \alpha_d}     }
\end{eqnarray}

From these expressions it can be seen that if $\alpha_s = \alpha_{s0} (T-T_s)$
and $\alpha_d = \alpha_{d0} (T-T_d) $  ($ T_d > T_s $) then the actual 
second order  phase transition is at $T_d$. At $T_s$ there is only a 
crossover, not accompanied by a phase transition. 
The coherence length $\xi_s$
doesn't diverge at $T_s$ but at a temperature much closer to $T_d$.

\subsubsection{Nonzero $|\Psi_s|$ at zero field}

This case can be considered naturally as a consequence of orthorhombic 
distortion of the lattice. Then the two representations  s and  $d_{x^2-y^2}$
can be mixed and  a bilinear term in $\Psi_d$ and $\Psi_s$ has to be included
in the free energy.
The calculations are presented in the Appendix. The main result is that
now the penetration depth acquires an anisotropy due to the nonzero value of
$\Psi_{s0}$. The important difference from the case with nonzero $\Psi_{s0}$
is the different temperature dependence which actually can distinguish
between the two cases.  The distortion of the order parameter is given by:

\begin{eqnarray}
|\Psi_d| = |\Psi_{d0}| + (\frac{H_0}{H_c})^2 \;(A_{1d} \; 
e^{-x/{\sqrt{2} \xi_d}}
+ A_{2d} \; e^{-2 x / \lambda} + A_{3d} \; e^{-x /{\sqrt{2} \xi_s}})  \\
|\Psi_s| = |\Psi_{s0}| + (\frac{H_0}{H_c})^2 \; (B_{1s} \; 
e^{-x/{\sqrt{2} \xi_s}}
+ B_{2s} \; e^{-2 x/  \lambda} + B_{3s} \; e^{-x /{\sqrt{2} \xi_d}} )            
\end{eqnarray}

The coefficients have been calculated in the Appendix and they depend
on the G.L. coefficients (terms beyond the first order in $\gamma_v$
have been neglected).  The thermodynamic
field $H_c$ has been taken as $ H_c = \frac{\hbar c}{\sqrt{2} e \xi_d \lambda}$.
For the $\hat{y}$-direction the results are the same with the substitution
$\gamma_v \rightarrow  - \gamma_v$.

Knowing the corrections of the order parameters, the effective penetration
depth can be computed easily (see Appendix). The main point of the calculation
is the temperature dependence of the anisotropy $\lambda_x/\lambda_y$.
In Fig. \ref{fig2} the temperature dependence of the ratio of 
relative corrections due to the field 
$\frac{\Delta\lambda_x/\lambda_x}{\Delta\lambda_y/\lambda_y}$ is plotted,
in the limit of strong type-II superconductivity ($\kappa >>1$).
This quantity is plotted in order to avoid the explicit field-dependence.
The relative corrections are larger in the temperature regime where the
anisotropy of the penetration depth at zero field is larger. The effect of
a small term which measures the orthorhombicity would be to get smooth
curves in the region close to $T_c$.

Again we see that the appearance of the s-wave admixture gives a strong 
signature in the penetration depth, this time in the nonlinear signal.

\section{DISCUSSION - CONCLUSIONS}
\label{sec:lever4}

The simple picture that this paper offers may be able to give a  
qualitative point of view of the several features that are observed 
simultaneously in penetration depth experiments. 
The basic conclusions are : \\ 
(i) The anisotropy in the penetration depth may
arise either as a consequence of directional-dependent ``masses'' or
as a consequence of the s and d-wave mixing.  As the figures show,
mixing may result in strong temperature and field dependence,
so the cases can be distinguished. In 
Ref.\ \cite{dynes}, considerable anisotropy was observed:
$\lambda_a > 1.5 \lambda_b$, but the temperature dependence was
not measured.  If the anisotropy is temperature dependent then the 
scenario with the non-zero $f_{s0}$ has to be investigated more 
carefully. It is ineresting that $\lambda_a/\lambda_b$
appeared to be strongly sample-dependent in these experiments.
While s-d mixing is expected to be a strong function of doping,
the effective mass ratio is not.  \\ 
(ii) The field - dependence of the penetration depth is of order ${H_0}^2$ 
for both cases at least in the geometry that it is described. This is in fact a 
consequence of the boundary conditions in the studied geometry.\\
(iii) The temperature dependence of the anisotropy can be derived explicitly 
and compared to experiments. Of course the G.L. equations are not valid at low 
temperatures, therefore nothing can be said for the experimentally observed 
crossover from T to $T^2$ dependence of the quantity ${1/\lambda(T)}^2$ 
\cite{hardy}. \\
(iv) Enhancement of the coupling and the transport along the c-axis if the
applied field is $in$ the $a-b$ plane. Then it's unavoidable to get screening
currents along the c-axis and actually the penetration depth has slightly
different values from the ones observed when $H_0 // \hat{z}$. \\
(v) Finally the question of d+s versus d-s states arises (and consequently
the equivalence of the two directions $\hat{x}$ and $\hat{y}$. In the model
that is described through the G.L. equations for a tetragonal crystal structure
a spontaneous symmetry breaking
is assumed that distinguishes between the two states and directions. The system
chooses one of the two equivalent states.  Measurements of the effects 
described here become much more complicated if domains of d+s and d-s form.  
In YBCO experiments, this requires detwinned crystals.

In Ref. \cite{sauls} the nonlinear Meissner effect was examined in the framework
of a microscopic model. The central mechanism that produces the nonlinearity
is the different contributions to the current from the ``co-moving'' (condensate
velocity parallel to Fermi velocity) and ``counter-moving'' (condensate velocity
antiparallel to Fermi velocity) excitations close to Fermi energy. Then in the
case of pure d-wave the correction is linear in the field. But even at very 
low temperatures this model predicts a transition to a regime where the 
corrections are of the order ${H_0}^2$. In addition to that a very small 
admixture of another component is able to change the behavior of the penetration
depth at very low temperatures as a function of temperature.  Our picture
is complementary to this in that it is nonly appropriate for
relatively high temperatures.  The effects predicted involve
relative values of the different gap components which are realistic.
The picture has the advantage of being rather general: 
note that effects such as impurity scattering
are included in the parameters of the G.L. functional.

There are several remarks on some other issues of the calculation 
of the penetration depth. The most important of these is the role of different
type of fluctuations. In Ref. \cite{moore} has been calculated the existence of
Off Diagonal Long Range Order (ODLRO) in the different ``states'' of the
superconductor. It is found that in the Meissner state this phase coherence 
is destroyed by phase fluctuations below two dimensions. Also only in strongly
type II superconductors, the fluctuations of the field can be considered 
unimportant. In Ref. \cite{tesanovic} the critical fluctuations in the 
penetration depth as the transition is approached from below is studied and it
is found that both the penetration depth and the coherence length diverge with
the same exponent which has value (0.53) very close to the mean-field one in 
three dimensions which was obtained in experimental work of Ref. \cite{lin} in
contrast with the value of $\sim 1/3$ obtained in Ref. \cite{kamal} on
$YBa_2Cu_3O_{6.95}$ pure bicrystals.
These issues suggest first that the calculations presented here are valid
in a region not very close to $T_c$ since in any case the small parameter of
the problem is in fact the applied field in comparison to thermodynamic field
which in turn is temperature dependent. Second the fluctuations in the case
of two order parameters will have to be included in a subsequent paper
because of the sensitivity of the second order parameter which is itself a
``secondary'' effect.

In summary we have presented the mean-field analysis in the Meissner state
of a superconductor in the case of one and two-component order parameter which
is relevant for the high-$T_c$ materials. In the latter case the picture of
the onset of the second component due to different reasons have been
presented and several predictions and qualitative explanations have been made.
 
\section{ACKNOWLEDGEMENTS}

We acknowledge helpful discussions with Andrey Chubukov,
Dirk Morr, and S.K. Yip. This work is supported by the NSF under the
Materials Research Science and Engineering Center Program,
Grant No. DMR-96-32527.

\section{APPENDIX}
We consider the case of the applied magnetic field $H_0$ parallel to the 
c-axis and varying along the $\hat{x}$ direction. The geometry is the same 
as in Part II. By following the same considerations as in Part II the phase 
can be written as $\phi=y \; g(x)$ and the gauge $\vec{A} = y \; a(x) \hat{x}$.
The magnitudes of the order parameters are denoted as $|\Psi_s| \equiv f_s$ and
$|\Psi_d| \equiv f_d $. We substitude all the above into the Eqs. 
(\ref{six}) and (\ref{seven}). Then by dividing the equations into real and 
imaginary parts and also by taking into account the power of y, 
we obtain the independent equations:
\begin{eqnarray}
\frac{dg}{dx} = \frac{2e}{c \hbar} a(x) 
\label{nine}\\
\gamma_d {\hbar}^2 (- \frac {d^2 f_d}{dx^2} + g^2 f_d) + \alpha_d f_d
+ \gamma_v {\hbar}^2 (\frac{d^2 f_s}{dx^2} + f_s g^2) + 2 \beta_2 {f_d}^3 
+ \beta_3 '{f_s}^2 f_d = 0 
\label{ten}\\
\gamma_s {\hbar}^2 (-\frac {d^2 f_s}{dx^2} + g^2 f_s) + \alpha_s f_s
- \gamma_v {\hbar}^2 (\frac {d^2 f_d}{dx^2} + g^2 f_d) + 
\beta_3 ' {f_d}^2 f_s =0
\label{eleven}
\end{eqnarray}

The magnetic field (which is explicitly contained in g) is considered as 
the small parameter of the problem, as in the case of the single order
parameter and we seek solutions of the form:
\begin{eqnarray}
g =  H_0  \; \tilde{g} \\
f_d = f_{d0} + {H_0}^2 \; f_{d1} \\
f_s = f_{s0} + {H_0}^2 \; f_{s1}
\end{eqnarray}

An important point here is that the dependence of $f_d$ and $f_s$ on the field
is quadratic because of the boundary conditions at $x=0$. With different 
boundary conditions, linear dependence on the field can be obtained. But this 
is not the case here. 
Finally, by gathering terms with the same power in the applied field,
the equations for $f_{d0}$, $f_{d1}$, $f_{s0}$, $f_{s1}$ are:
\begin{eqnarray}
\alpha_d f_{d0} + 2 \beta_2 {f_{d0}}^3 = 0 
\label{twelve}\\
\alpha_s f_{s0} + 2 \beta_1 {f_{s0}}^3 = 0 
\label{thirteen}\\
\gamma_d {\hbar}^2 (- \frac{d^2 f_{d1}}{dx^2} + {\tilde{g}}^2 f_{d0})
+ \alpha_d f_{d1} + \gamma_v {\hbar}^2 ( \frac{d^2 f_{s1}}{dx^2} +
{\tilde{g}}^2 f_{s0} ) + 6 \beta_2 {f_{d0}}^2 f_{d1} + \beta_3 '
{f_{s0}}^2 f_{d1} = 0 
\label{fourteen}\\
\gamma_s {\hbar}^2 (- \frac{d^2 f_{s1}}{dx^2} + {\tilde{g}}^2 f_{s0}) +
\alpha_s f_{s1} - \gamma_v {\hbar}^2 (\frac{d^2 f_{d1}}{dx^2} + 
{\tilde{g}}^2 f_{d0} ) + \beta_3 ' {f_{d0}}^2 f_{s1} = 0
\label{fifteen}
\end{eqnarray}

The possible solutions of the Eq. (\ref{thirteen}) are $f_s =0 $ and
$f_s = \sqrt{-\frac{\alpha_s}{2 \beta_1}}$. In the results presented in the
text the difference in the two cases is discussed. The calculations
here are presented under the hypothesis of a nonzero $f_s$.
From Eq. (\ref{twelve}, \ref{thirteen}) the zeroth order value of the two 
order parameters can be easily obtained.
Then the equations (\ref{fourteen}, \ref{fifteen}) suggest the following :
It is natural to expect dependence of $f_{d1}$ on the coherence length
$\xi_d$ as well as the penetration depth $\lambda$ (exactly like the case of 
one order parameter) but due to the appearance of the terms which depend on 
$f_s$ it is also expected a small dependence on the coherence length $\xi_s$. 
The same, by symmetry, is true for the correction of the s-component $f_{s1}$.
So, the homogeneous parts of Eq. (\ref{fourteen}, \ref{fifteen}) have  solutions
which depend on the corresponding coherence length of each component. Then 
the non-homogeneous differential equations mix the dependences as discussed 
above. For the exact form of the solutions, the boundary conditions have to 
be taken into account. They take the explicit form : $ f_{s1}(x=\infty)= f_{d1}
(x=\infty)= 0$, $f_s'(x=0) = f_d'(x=0)=0$ and
$a(x=0) = -H_0$, $a(x=\infty)=0$.
Having elaborated the crucial points, it is straightforward,
after some lengthy algebra to obtain the full solutions-corrections to
the two order parameters.
The homogeneous parts of the equations accept solutions of the form :
$f_{d1,h}= A_{1d} \; exp(-x/{\sqrt{2}\xi_d})$ and $f_{s1,h}= B_{1s} \;
exp(-x/{\sqrt{2} \xi_s})$ . If we substitute these expressions into the
homogeneous parts of the differential equations, we obtain the expressions
for the two coherence lengths, which are the {\it same} as in the case of
zero $f_{s0}$ up to order $\gamma_v ^2$.
The solutions of the non-homogeneous parts are : $f_{d1,inh}=
 A_{2d} \; exp(-2x/\lambda) + A_{3d} \; exp(-x/{\sqrt{2} \xi_s})$ and
$f_{s1,inh}= B_{2s} \; exp(-2x/\lambda) + B_{3s} \; exp(-x/{\sqrt{2} \xi_d})$.
The coefficients can be evaluated by substituting the above expressions into
the inhomogeneous parts of the differential equations and by using the
boundary conditions. Finally  we get :

\begin{eqnarray}
\nonumber
A_{1d} &=& \frac{4 \sqrt{2}}{\xi_d} [-\frac{-\gamma_d f_{d0} (-4/{\lambda}^2 +
1/{2{\xi_s}^2}) + \gamma_v f_{s0} (8/{\lambda}^2 - 1/{2{\xi_s}^2})}
{\lambda \gamma_d (-4/{\lambda}^2 + 1/{2{\xi_d}^2})(-4/{\lambda}^2 +
1/{2{\xi_s}^2})} \\
&+& \frac{\lambda \gamma_v f_{s0}}{ {\xi_s}^2 \gamma_d
(1- (\xi_s/\xi_d)^2) (4/{\lambda}^2 - 1/{2{\xi_s}^2}) }] \\
A_{2d}&=&2 \frac{ -\gamma_d f_{d0} (-4/{\lambda}^2 + 1/{2{\xi_s}^2})+
\gamma_v f_{s0} (8/{\lambda}^2 - 1/{2{\xi_s}^2})}{  {\xi_d}^2 \gamma_d
(-4/{\lambda}^2 + 1/{2{\xi_d}^2})(-4/{\lambda}^2 + 1/ {2{\xi_s}^2})}\\
A_{3d} &=& 4\sqrt{2} \frac{\gamma_v \lambda f_{s0} }{\gamma_d \xi_s {\xi_d}^2
(1 - (\xi_s/\xi_d)^2) (-4/ {\lambda}^2 + 1/{2{\xi_s}^2})} \\
\nonumber
B_{1s}&=& 4 \sqrt{2}/ {\xi_d}^2 [-\frac{\lambda (-\gamma_s f_{s0}
(-4/\lambda^2 + 1/{2{\xi_d}^2}) + \gamma_v f_{d0}
(8/\lambda^2 - 1/{2{\xi_d}^2}))}
{\xi_s \gamma_s (-4/\lambda^2 + 
1/{2{\xi_d}^2})(-4/\lambda^2 + 1/{2{\xi_s}^2})} \\
&+& \frac{\xi_s \gamma_v f_{d0}}{\lambda \gamma_s (1 - (\xi_d/ \xi_s)^2)
(4/\lambda^2 - 1/{2{\xi_d}^2} )} ] \\
B_{2s} &=& 2 \frac{- \gamma_s f_{s0} (-4/\lambda^2 + 1/{2{\xi_d}^2}) +
\gamma_v f_{d0} (8/\lambda^2 - 1/{2{\xi_d}^2})}{{\xi_d}^2 \gamma_s
(-4/\lambda^2 + 1/{2{\xi_d}^2})(-4/\lambda^2 + 1/ {2{\xi_s}^2})}  \\
B_{3s} &=& 4\sqrt{2} \frac{\gamma_v  f_{d0} }{\gamma_d \xi_s \xi_d
(1 - (\xi_s/\xi_d)^2) (-4/ \lambda^2 + 1/{2{\xi_s}^2}) }
\end{eqnarray}

The effective penetration depth then can be obtained from the solution of the 
Eq. (\ref{eight})  with the appropriate boundary conditions that have been 
described, taking into account Eq. (\ref{nine}). The calculation is 
straightforward and the result is :

\begin{eqnarray}
\lambda_{eff.}= \lambda (1 + (\frac{H_0}{H_c})^2 \; ( \frac{2 C_1}{\lambda/\xi_d
+ 2 \sqrt{2}} + \frac{C_2}{4} + \frac{2 C_3}{\lambda/\xi_s + 2 \sqrt{2} } )
\end{eqnarray}
 
The coefficients $C_i$, i=1,2,3 above are given by :

\begin{eqnarray}
\nonumber
C_1 = \frac{E}{D} \; A_{1d}  + \frac{F}{D}  \; B_{3s} \\
\nonumber
C_2 = \frac{E}{D} \; A_{2d}  + \frac{F }{D} \; B_{2s} \\
\nonumber
C_3 = \frac{E}{D} \; A_{3d}  + \frac{F}{D}  \; B_{1s}
\end{eqnarray}

where $D = \gamma_d {f_{d0}}^2 + \gamma_s {f_{s0}}^2 + \gamma_v f_{s0} f_{d0}$,
 $E= 2 \gamma_d f_{d0} + \gamma_v f_{s0} $ and $F = 2 \gamma_s f_{s0} + \gamma_v
f_{d0} $.

\begin{figure}
\caption{ The anisotropy $\lambda_x / \lambda_y$ at zero field as a function
of temperature. The values of the parameters that have been used are 
$|\Psi_s|/ |\Psi_d| = 0.1 \; (\frac{t-0.8}{t-1})^{1/2}$, 
$\gamma_s/ \gamma_d =1 $ and $\gamma_v = 0.6 \gamma_d$.   }
\label{fig1}
\end{figure}

\begin{figure}
\caption{ The ratio of the relative corrections due to the applied field in the
regime of strong type-II superconductor. The values of the parameters are the 
same as in Fig. 1. In addition to that $\beta_1 = \beta_2 = \beta_3' /2$ }
\label{fig2}
\end{figure}

\end{document}